\documentclass[3p,times]{elsarticle}

\usepackage{ecrc}

\volume{00}

\firstpage{1}

\journalname{Nuclear Physics B Proceedings Supplement}

\runauth{}

\jid{nuphbp}

\jnltitlelogo{Nuclear Physics B Proceedings Supplement}

\usepackage{amssymb}

\usepackage[figuresright]{rotating}

\usepackage{color}
\usepackage{amsmath} 
\usepackage{amssymb} 
\usepackage[english]{babel}
\usepackage{graphics}
\usepackage{xspace}
\usepackage[gen]{eurosym}
\usepackage[font=footnotesize,format=plain,labelfont=bf,up,textfont=it,up]{caption}
\usepackage{graphics}
\usepackage{subfig}
\usepackage{latexsym}
\usepackage[mathscr]{eucal} 
\usepackage{latexsym}
\usepackage[left]{lineno}

\newcommand{\nue}{\ensuremath{\nu_{e}}\xspace}

\newcommand{\numu}{\ensuremath{\nu_{\mu}}\xspace}

\newcommand{\nubarmu}{\ensuremath{\overline{\nu}_{\mu}}\xspace}

\newcommand{\numunust}{\ensuremath{\nu_\mu \rightarrow \nu_s}\xspace}

\newcommand{\boss}[2]{\ensuremath{\rlap{\kern-2.5pt\ensuremath{\overset{\scriptscriptstyle(-)}{\phantom{#1}}}}{\ensuremath{{#1}_{#2}}}}}

\newcommand{\BARII}                   {1}
\newcommand{\BARIU}                 {2}
\newcommand{\BOLOGNAI}         {3}
\newcommand{\BOLOGNAU}       {4}
\newcommand{\CERN}                  {5}
\newcommand{\FRASCATI}          {6}
\newcommand{\LECCEI}             {7}
\newcommand{\LECCEU}           {8}
\newcommand{\LECCEING}       {9}
\newcommand{\LEBEDEV}          {10}
\newcommand{\MSU}                 {11}
\newcommand{\PADOVAI}          {12}
\newcommand{\PADOVAU}        {13}
\newcommand{\ROMAU}            {14}
\newcommand{\ZAGREB}             {15}
\newcommand{\NessieInstitutes}{
\BARII        . INFN, Sezione di Bari, 70126 Bari, Italy \\
\BARIU        . Dipartimento di Fisica dell'Universit\`a  di Bari, 70126 Bari, Italy \\
\BOLOGNAI     . INFN, Sezione di Bologna, 40127 Bologna, Italy \\
\BOLOGNAU     . Dipartimento di Fisica e Astronomia dell'Universit\`a  di Bologna, 40127 Bologna, Italy \\
\CERN             .  European Organization for Nuclear Research (CERN), Geneva, Switzerland \\
\FRASCATI    . Laboratori Nazionali di Frascati dell'INFN, 00044 Frascati (Roma), Italy \\
\LECCEI        . INFN, Sezione di Lecce, 73100 Lecce, Italy \\
\LECCEU        . Dipartimento di Matematica e Fisica dell'Universit\`a  del Salento, 73100 Lecce, Italy \\
\LECCEING        . Dipartimento di Ingegneria dell'Innovazione dell'Universit\`a  del Salento, 73100 Lecce, Italy \\
\LEBEDEV     . Lebedev Physical Institute of Russian Academy of Science, Leninskie pr., 53, 119333 Moscow, Russia.\\
\MSU       . Lomonosov Moscow State University (MSU SINP), 1(2) Leninskie gory, GSP-1, 119991 Moscow, Russia\\
\PADOVAI      . INFN, Sezione di Padova, 35131 Padova, Italy \\
\PADOVAU      . Dipartimento di Fisica e Astronomia dell'Universit\`a  di Padova, 35131 Padova, Italy \\
\ROMAU        . Dipartimento di Fisica dell'Universit\`a  di Roma ``La Sapienza" and INFN, 00185 Roma, Italy \\
\ZAGREB             . Rudjer Boskovic Institute, Bijenicka 54, 10002 Zagreb, Croatia\\
\ddag~Also at Centre de Recherche en Astronomie Astrophysique et Geophysique, Alger, Algeria\\
}
\newcommand{\NessieAuthorList}{
\noindent 
A.~Anokhina$^{\MSU}$,
A.~Bagulya$^{\LEBEDEV}$,
M.~Benettoni$^{\PADOVAI}$,
P.~Bernardini$^{\LECCEU, \LECCEI}$,
R.~Brugnera$^{\PADOVAU, \PADOVAI}$,
M.~Calabrese$^{\LECCEI}$,
A.~Cecchetti$^{\FRASCATI}$,
S.~Cecchini$^{\BOLOGNAI}$,
M.~Chernyavskiy$^{\LEBEDEV}$,
P.~Creti$^{\LECCEI}$,
F.~Dal~Corso$^{\PADOVAI}$,
O.~Dalkarov$^{\LEBEDEV}$,
A.~Del~Prete$^{\LECCEING, \LECCEI}$,
G.~De~Robertis$^{\BARII}$,
M.~De~Serio$^{\BARIU, \BARII}$,
L.~Degli~Esposti$^{\BOLOGNAI}$,
D.~Di~Ferdinando$^{\BOLOGNAI}$,
S.~Dusini$^{\PADOVAI}$,
T.~Dzhatdoev$^{\MSU}$,
C.~Fanin$^{\PADOVAI}$,
R.~A.~Fini$^{\BARII}$,
G.~Fiore$^{\LECCEI}$,
A.~Garfagnini$^{\PADOVAU, \PADOVAI}$,
S.~Golovanov$^{\LEBEDEV}$,
M.~Guerzoni$^{\BOLOGNAI}$,
B.~Klicek$^{\ZAGREB}$,
U.~Kose$^{\CERN}$,
K.~Jakovcic$^{\ZAGREB}$,
G.~Laurenti$^{\BOLOGNAI}$,
I.~Lippi$^{\PADOVAI}$,
F.~Loddo$^{\BARII}$,
A.~Longhin$^{\FRASCATI}$,
M.~Malenica$^{\ZAGREB}$,
G.~Mancarella$^{\LECCEU, \LECCEI}$,
G.~Mandrioli$^{\BOLOGNAI}$,
A.~Margiotta$^{\BOLOGNAU, \BOLOGNAI}$,
G.~Marsella$^{\LECCEU, \LECCEI}$,
N.~Mauri$^{\FRASCATI}$,
E.~Medinaceli$^{\PADOVAU, \PADOVAI}$,
A.~Mengucci$^{\FRASCATI}$,
R.~Mingazheva$^{\LEBEDEV}$,
O.~Morgunova$^{\MSU}$,
M.~T.~Muciaccia$^{\BARIU, \BARII}$,
M.~Nessi$^{\CERN}$,
D.~Orecchini$^{\FRASCATI}$,
A.~Paoloni$^{\FRASCATI}$,
G.~Papadia$^{\LECCEING, \LECCEI}$,
L.~Paparella$^{\BARIU, \BARII}$,
L.~Pasqualini$^{\BOLOGNAU,\BOLOGNAI}$,
A.~Pastore$^{\BARII}$,
L.~Patrizii$^{\BOLOGNAI}$,
N.~Polukhina$^{\LEBEDEV}$,
M.~Pozzato$^{\BOLOGNAU, \BOLOGNAI}$,
M.~Roda$^{\PADOVAU, \PADOVAI}$,
T.~Roganova$^{\MSU}$,
G.~Rosa$^{\ROMAU}$,
Z.~Sahnoun$^{\BOLOGNAI \ddag}$,
S.~Simone$^{\BARIU, \BARII}$,
C.~Sirignano$^{\PADOVAU, \PADOVAI}$,
G.~Sirri$^{\BOLOGNAI}$,
M.~Spurio$^{\BOLOGNAU, \BOLOGNAI}$,
L.~Stanco$^{\PADOVAI, a}$,
N.~Starkov$^{\LEBEDEV}$,
M.~Stipcevic$^{\ZAGREB}$,
A.~Surdo$^{\LECCEI}$,
M.~Tenti$^{\BOLOGNAU, \BOLOGNAI}$,
V.~Togo$^{\BOLOGNAI}$,
M.~Ventura$^{\FRASCATI}$ and
M.~Vladymyrov$^{\LEBEDEV}$.\\
{\em (a)} Spokesperson
}

\begin{document}

\begin{frontmatter}



\dochead{}

\title{The NESSiE way to searches for sterile neutrinos at FNAL}


\author{L. Stanco, for the NESSiE Collaboration}

\address{INFN-Padova, Va Marzolo, 8 I-35131 Padova, Italy}

\begin{abstract}
\noindent Neutrino physics is nowadays receiving more and more attention as a possible source of information for the long--standing problem
of new physics beyond the Standard Model. The recent measurement of the mixing angle $\theta_{13}$ in the standard 
mixing oscillation scenario encourages us to pursue the still missing results on leptonic CP violation and absolute neutrino
masses. However, puzzling measurements exist that deserve an exhaustive evaluation.

The NESSiE Collaboration has been setup to undertake conclusive experiments to clarify the  {\em muon--neutrino disappearance} measurements at small $L/E$,
which will be able to put severe constraints to models with more than the three-standard neutrinos, or even to robustly measure 
the presence of a new kind of neutrino oscillation for the first time.
To this aim the use of the current FNAL--Booster neutrino beam for a Short--Baseline experiment
has been carefully evaluated. Its recent proposal  refers to the use of magnetic spectrometers at two different sites, Near and Far ones.
Their positions have been extensively studied, together with the possible performances of two OPERA--like spectrometers.
The proposal is constrained by availability of existing hardware and a time--schedule compatible with the undergoing project of a multi--site Liquid--Argon detectors at FNAL.

The experiment to be possibly setup at Booster  will allow to definitively clarify the current $\numu$ disappearance tension with $\nue$ appearance and disappearance at the eV mass scale.
\end{abstract}


\end{frontmatter}

\vskip5cm
{\em \noindent ICHEP2014, \\
		2--9 July 2014,\\
		Valencia, Spain}


\newpage
\centerline{\bf The NESSiE Collaboration for the FNAL experiment}

{\noindent \\ \NessieAuthorList }

\begin{flushleft}
\footnotesize\em{\NessieInstitutes }
\end{flushleft}

\newpage

\section{Introduction and Physics Overview}
\label{intro}

The unfolding of the physics of the neutrino is a long and exciting history spanning the last 80 years. Over this time the interchange of 
theoretical hypotheses and experimental facts has been one of the most fruitful demonstrations of the progress of knowledge in physics.
The work of the last decade and a half finally brought a coherent picture within the Standard Model (SM) (or some small extensions of it),
namely the mixing of three neutrino flavour states with three  $\nu_1$, $\nu_2$ and $\nu_3$ mass eigenstates. 
The last unknown mixing angle, $\theta_{13}$, was recently measured~\cite{theta13} but still
many questions remain unanswered to completely settle the scenario: the absolute masses, 
the Majorana/Dirac nature and the existence and magnitude of leptonic CP violation.
Answers to these questions will beautifully complete the (standard) three--neutrino model but they will hardly provide an insight into new physics 
Beyond the Standard Model (BSM). 
Many relevant questions will stay open: the reason for the characteristic nature of neutrinos, the relation between the
leptonic and hadronic sectors of the SM, the origin of Dark Matter and, overall, where and how to look for BSM physics.
Neutrinos may be an excellent source of BSM physics and their history is supporting that at length.

There are actually several experimental hints for deviations from the ``coherent'' picture described above.
Many unexpected results, not statistically significant on a single basis, appeared also in the last decade and a half,
bringing attention to the hypothesis of the existence of {\em sterile neutrinos}~\cite{pontecorvo}. A White Paper~\cite{whitepaper},
contains a comprehensive review of these issues. 
In particular we would like to focus on tensions in many phenomenological models that grew up with experimental results on 
neutrino/antineutrino oscillations at Short--Baseline (SBL) and with the more recent, carefully recomputed, antineutrino fluxes from 
nuclear reactors. The main source of tension corresponds to the lack so far of any \numu disappearance signal~\cite{kopp-tension}.
This tension has been strengthened by the recent exclusion limits reported at the NEUTRINO2014 conference~\cite{recent-limits}.

This scenario promoted several proposals for new, exhaustive evaluations of the neutrino behaviour
at SBL. Since the end of 2012 CERN is undergoing a study to setup a Neutrino Platform, with a new
infrastructure at the North Area that, for the time being, will not include a new neutrino beam~\cite{edms}. Meanwhile FNAL is
welcoming proposals of experiments to exploit the physics potentials of their two existing neutrino beams, the Booster and the NUMI beams, following the
recent recommendations from USA HEP-P5 report~\cite{p5}.
Two recent proposals~\cite{LAr1-ND, ICARUSFNAL} have been submitted for experiments of SBL at the Booster beam, to complement
the about to start MicroBooNE experiment~\cite{microboone}. They 
are both based on the Liquid Argon technology and aim to measure the \nue appearance at SBL, with possibilities to 
study the \numu disappearance. Possible use of magnetic spectrometers at two different sites at FNAL--Booster beam have been proposed by the
NESSiE collaboration and discussed in detail in~\cite{nessie-fnal}.

\begin{figure}[htbp]
\centering
\includegraphics[width=10cm]{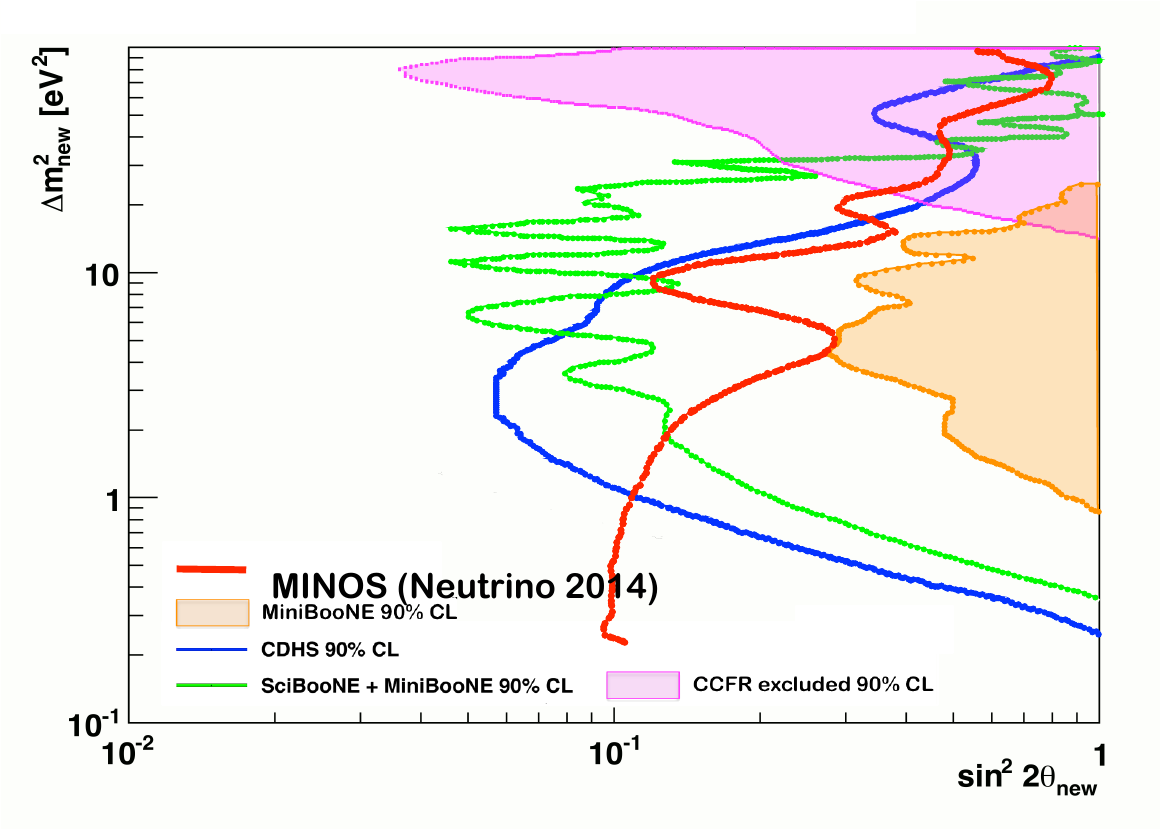}
\caption{The current exclusion limits on the \numu disappearance searches at the eV$^2$ scale.
Blue (green) line: old (recent) exclusion limits on \numu from previous CDHS~\cite{CDHS} and recent MiniBooNE/SciBooNE~\cite{mini-sci-mu} measurements.
The two filled areas correspond to the exclusion limits on the \nubarmu from CCFR~\cite{ccfr} and MiniBooNE--alone~\cite{mini-mu} experiments (at 90\% C.L.).
The red curve corresponds to the very recent result from MINOS~\cite{recent-limits}.}
\label{fig:old-res}
\end{figure}

The NESSiE proposal is based on the following considerations:
\begin{itemize}
\item the measurement of \numu spectrum and trend is mandatory for a correct interpretation of the \nue data, even in case of 
a null result for the latter;
\item a decoupled measurement of \nue and \numu interactions will allow to reach in the analyses the percent--level systematics due to the 
different cross--sections;
\item very massive detectors are mandatory to collect a large number of events thus improving the disentangling
of systematic effects.
\item limited experimental data are available on \numu disappearance at SBL: the dated CDHS experiment~\cite{CDHS} and
the more recent results from MiniBooNE~\cite{mini-mu}, a joint MiniBooNE/SciBooNE analysis~\cite{mini-sci-mu} and
MINOS and SK~\cite{recent-limits}. 
The latter results slightly extend the $\nu_\mu$ disappearance exclusion region set by CDHS.
Fig.~\ref{fig:old-res} shows the excluded regions in the space parameters for the \numunust oscillation, obtained through
\numu disappearance experiments. The mixing angle is denoted as $\theta_{new}$ and the squared mass difference as
$\Delta m^2_{new}$. The region with $\sin^2(2\theta_{new})<0.1$ is largely still unconstrained.
\end{itemize}

\section{Proposal for the FNAL--Booster}\label{expo}

Motivated by the present scenario a detailed study of the physics case for the FNAL--Booster beam was performed. 
The study follows the similar analysis developed for the CERN--PS and CERN--SPS cases~\cite{nessie,larnessie} and the study in~\cite{stancoetal}.
We pondered many detector configurations investigating experimental aspects not fully addressed by the LAr detection. This includes 
the measurements of the lepton charge on event--by--event basis and its energy over a wide range.
Indeed, muons from Charged Current (CC) neutrino interactions play an important role in disentangling 
different phenomenological scenarios provided their charge state is determined. Also,
the study of muon appearance/disappearance can benefit from the large statistics of muonic--CC events from the primary 
neutrino beam. In the FNAL--Booster beam the antineutrino contribution is rather small and it then becomes a systematic
effect to be taken into account.

Results of our study are reported in detail in the full NESSiE proposal~\cite{nessie-fnal}. 
We aim to design, construct and install two spectrometers at two sites,
``Near'' (at 110 m, on--axis) and ``Far'' (at 710 m, on surface), in line with
the FNAL--Booster, fully compatible with the proposed LAr detectors. 
Profiting of the large mass of the two spectrometer--systems, their stand--alone performances are exploited for the \numu 
disappearance study. Besides, complementary measurements with LAr can be undertaken to increase their control of
systematic errors.

Some important practical constraints were assumed in order to draft a proposal on a
conservative, manageable basis, with sustainable timescale and cost-wise.
Well known technologies were considered as well as re--using large parts of existing detectors. 

The momentum and charge state measurements of muons in a wide range, from few hundreds MeV/c to several 
GeV/c, over a $> \ 50\ {\rm m}^2$ surface, is an extremely challenging task if constrained by an order of 1 million~\euro\ budget
for construction and installation.  Running costs must be kept at low level, too.



We believe to have succeeded in developing a substantial proposal that, by keeping the systematic error at the level of $1\div 2$\% for the measurements of the 
\numu interactions, will allow to:
\begin{itemize}
	\item
measure \numu disappearance in the almost entire available momentum range ($p_{\mu}\ge 500$~MeV/c). The capability in 
rejecting/observing the anomalies over the whole expected parameter space of sterile neutrino oscillations is a key feature, since the momentum
range corresponds to an almost equivalent $\Delta m^2_{new}$ interval;
\item collect a very large statistical sample in order to span the oscillation mixing parameter down to till un--explored regions ($\sin^2(2\theta_{new})\gtrsim 0.01$);
	\item
measure the neutrino flux at the Near detector, in the relevant muon momentum range, to keep the systematic errors at the lowest possible values;
\item measure the sign of the muon charge to
separate \numu from \nubarmu for systematics control.
\end{itemize}

In the following the key points of the proposal are recalled.

\subsection{The Booster Neutrino Beam (BNB)}

The neutrino beam~\cite{G4BNBflux} is produced using protons with a
kinetic energy of 8 GeV extracted from the Booster and directed to a
Beryllium cylindrical target with a length of 71~cm and 1~cm diameter.  
The target is surrounded by a magnetic focusing horn pulsed
with a 170~kA current at a rate of 5~Hz. Secondary mesons are
projected into a 50~m long decay pipe where they are allowed to decay
in flight before reaching by an absorber and the ground
material.  An additional absorber can be placed in the decay pipe at
about 25~m from the target\footnote{\noindent This configuration,
  which is not currently in use, could eventually alter the beam
  properties (i.e. providing a more point--like source for the Near
  site) thus allowing for extra experimental constraints on the
  systematic errors.}.  Neutrinos travel about horizontally at a depth
of about 7~m underground.

A booster acceleration cycle typically contains about $4.5\times 10^{12}$
protons. Batches have a duration of 1.6 $\mu$s and are subdivided
into 84 bunches, about 4~ns wide and spaced by about
19~ns. The rate of batch extraction is limited by the horn pulsing at
5~Hz. This timing structure provides a very powerful handle to constraint
background from cosmic rays.

\subsection{The Far--to--Near ratio (FNR)}

The uncertainty on the absolute $\nu_\mu$ flux at MiniBooNE stays below
20\% for energies below 1.5~GeV while it increases drastically above that energy.
The uncertainty is dominated by the knowledge of hadronic interactions
of protons on the Be target, which affects the angular and momentum
spectra of neutrino parents emerging from the target. The obtained result~\cite{G4BNBflux}
is based on experimental data obtained by the HARP
and E910 collaborations.

Such a large uncertainty makes the use of two or more identical
detectors at different baselines mandatory when searching for small
disappearance phenomena. The ratio of the event rates at the Far and
Near detectors (FNR) as function of neutrino energy is a convenient
variable since it benefits at first order from cancellation of common
proton--target and neutrino cross--sections systematics and of the effects of reconstruction efficiencies.

Thanks to these cancellations the uncertainty on the FNR or,
equivalently, on the Far spectrum extrapolated from the Near spectrum
is usually at the percent level ranging in the 0.5--5.0\% interval.

It can be noted that, even in the absence of oscillations, the energy
spectra in the two detectors are different, thus leading to a non--flat
FNR. This is especially true if the Near detector is at a distance
comparable to the length of the decay pipe. It is therefore essential
to master the knowledge of the FNR for physics searches.

Compared to the Far site the solid angle subtended by the Near detector is larger. Moreover
neutrinos originating from meson decays at the end of the decay pipe
have a larger probability of being detected.  On
the contrary, only neutrinos produced in a narrow forward cone
will cross the Far detector.

\begin{figure}[htbp]
\centering
\includegraphics[scale=0.42,type=pdf,ext=.pdf,read=.pdf]{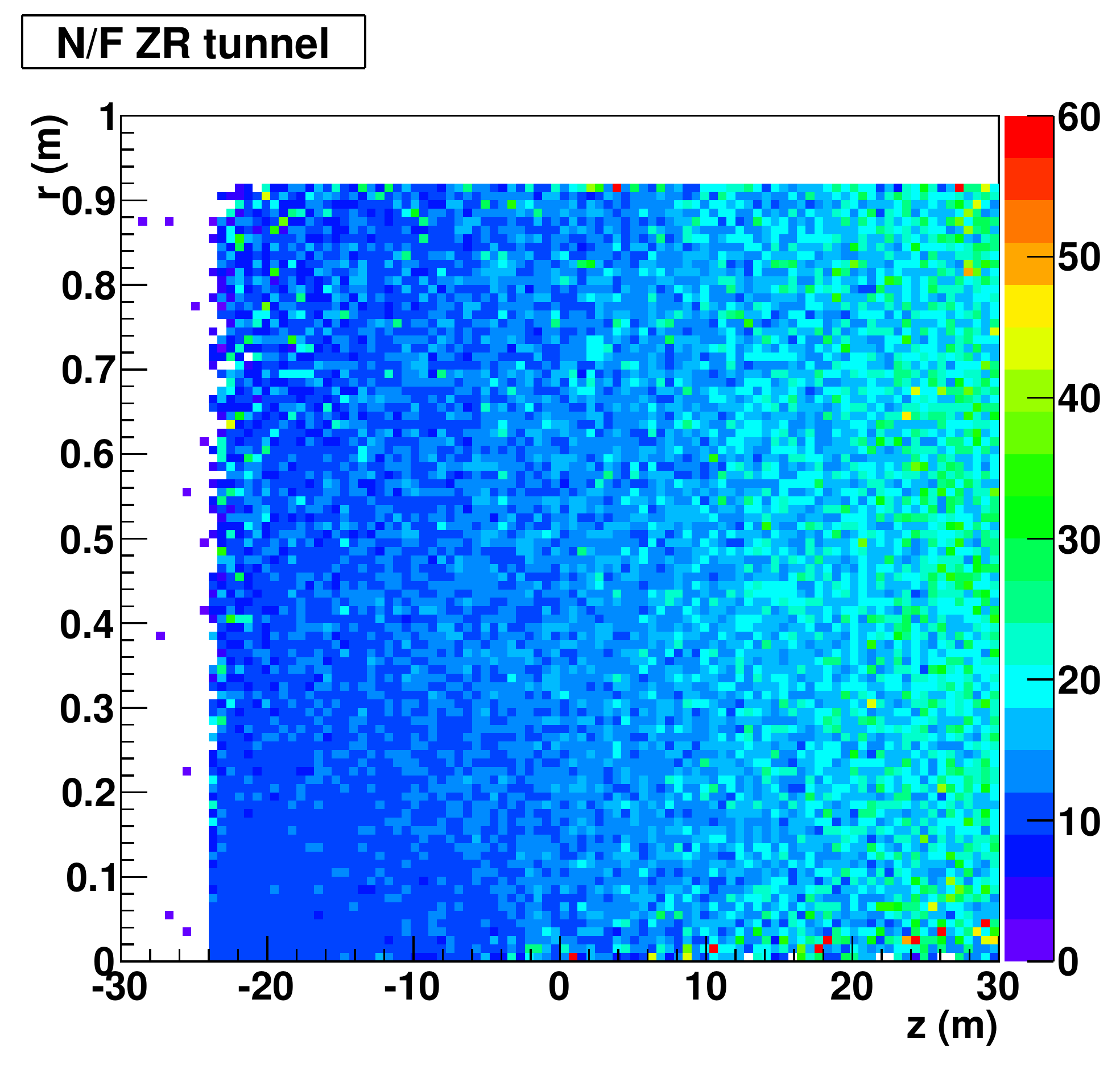}
\caption{Ratio between the $Z$--$R$ distributions of neutrino
  production points for neutrinos observed in a Near detector (at 110 m) over
  neutrinos observed in a Far detector (at 710 m).  Two effects are most relevant: there is no apparent dependence on the
  radial $R$ distribution; and, as expected, the Near detector has a
  higher acceptance for neutrinos produced in the most downstream part
  of the decay pipe, i.e. at high $Z$.}
\label{fig:teff}
\end{figure}

Assuming realistic detector sizes (see Table~\ref{tab:NFD}) the effect of the increased acceptance of the Near detector for
neutrinos from late decays
is illustrated in Fig.~\ref{fig:teff}. The ratio of the
distributions of the neutrino production points (radius $R$ vs
longitudinal coordinate $Z$) is shown for a sample crossing a Near and a Far
detector placed at 110 and 710 m from the target, respectively.
Neutrinos produced at large $Z$ can be detected in the Near detector even if 
produced at relatively large angles, thus enhancing the low energy
part of the spectrum.  On the other hand neutrinos coming from meson decays  late in
the decay pipe are originating from the fast pion component which is more 
forward--boosted. The former effect is the leading one so the net effect is a
softer neutrino energy spectrum at the Near site.

From these qualitative considerations it becomes clear that the
prediction of the FNR is a delicate task requiring a full simulation
of the neutrino beamline and of the detector acceptance, and a careful control of the
systematic uncertainties.

All the contributions to the systematic uncertainties have been
studied in detail by the MiniBooNE collaboration in~\cite{G4BNBflux}. 
The dominant contribution comes from
the knowledge of the hadroproduction double differential ($p$,
$\theta$) cross--sections in 8 GeV $p$--Be interactions.
At first order these contributions factorize out using a double site.

In order to understand how the hadroproduction uncertainty affects the accuracy on
FNR for the specific case of our experiment we 
developed anew the beamline simulation.
The angular and momentum distribution of pions exiting the Be target were studied using:
\begin{itemize}
\item FLUKA 2011.2b~\cite{fluka},
\item GEANT4 (v4.9.4 p02, QGSP 3.4 physics list),
\item a Sanford--Wang parametrization determined from a fit of the HARP and E910 data sets in~\cite{G4BNBflux},
\end{itemize}
\begin{equation}
\frac{d^2\sigma}{dpd\Omega} = c_1 p^{c_2}\left(1-\frac{p}{p_B-1}\right)\exp\left(-\frac{p^{c_3}}{p_B^{c_4}}-c_5\theta(p-c_6p_B\cos^{c_7}\theta)\right)
\label{eq:SW}
\end{equation}
where $p$, $\theta$ are the angular and momentum distribution of pions exiting the Be target
while $p_B$ is the proton beam momentum in GeV/c.

\subsection{The experimental sites}

A set of six configurations were studied considering a
combination of distances (110, 460 and 710 m), on--axis or off--axis
configurations and different fiducial sizes of the detectors.  Their
geometrical parameters are given in Tab.~\ref{tab:confs}. 
The FNRs for the six considered configurations using either FLUKA,
GEANT4 or the Sanford--Wang parametrization for the simulation of
$p$--Be interactions have been studied via several
Monte Carlo samples.  

\begin{table}
\scriptsize
\centering
\begin{tabular}{|c|c|c|c|c|c|c|}
\hline
config. &$L_N$ (m)&$L_F$ (m)&$y_N$ (m)& $y_F$ (m)& $s_N$ (m)& $s_F$ (m)\\
\hline
1 &110&710&0& 0 &4  &8 \\
\hline
2 &110&710&0& 0 &1.25  &8 \\
\hline
3 &110&710&1.4& 11 & 4 & 8 \\
\hline
4 &110&710&1.4& 11 &1.25  &8 \\
\hline
5 &460&710&7& 11 & 4 & 8 \\
\hline
6 &460&710&6.5& 10 & 4 & 6 \\
\hline
\end{tabular}
\caption{Near--Far detectors configurations. $L_{N(F)}$ is the distance of the Near (Far) detector
from the target. $y_{N(F)}$ is the vertical coordinate of the center of the fiducial area of the Near (Far) detector with respect to the beam axis
which lies at about -7~m underneath the ground surface. 
$s_{N(F)}$ is the dimension of the fiducial area of the Near (Far) detector.}
\label{tab:confs}
\end{table}

\begin{figure}
\centering
\includegraphics[scale=0.50,type=pdf,ext=.pdf,read=.pdf]{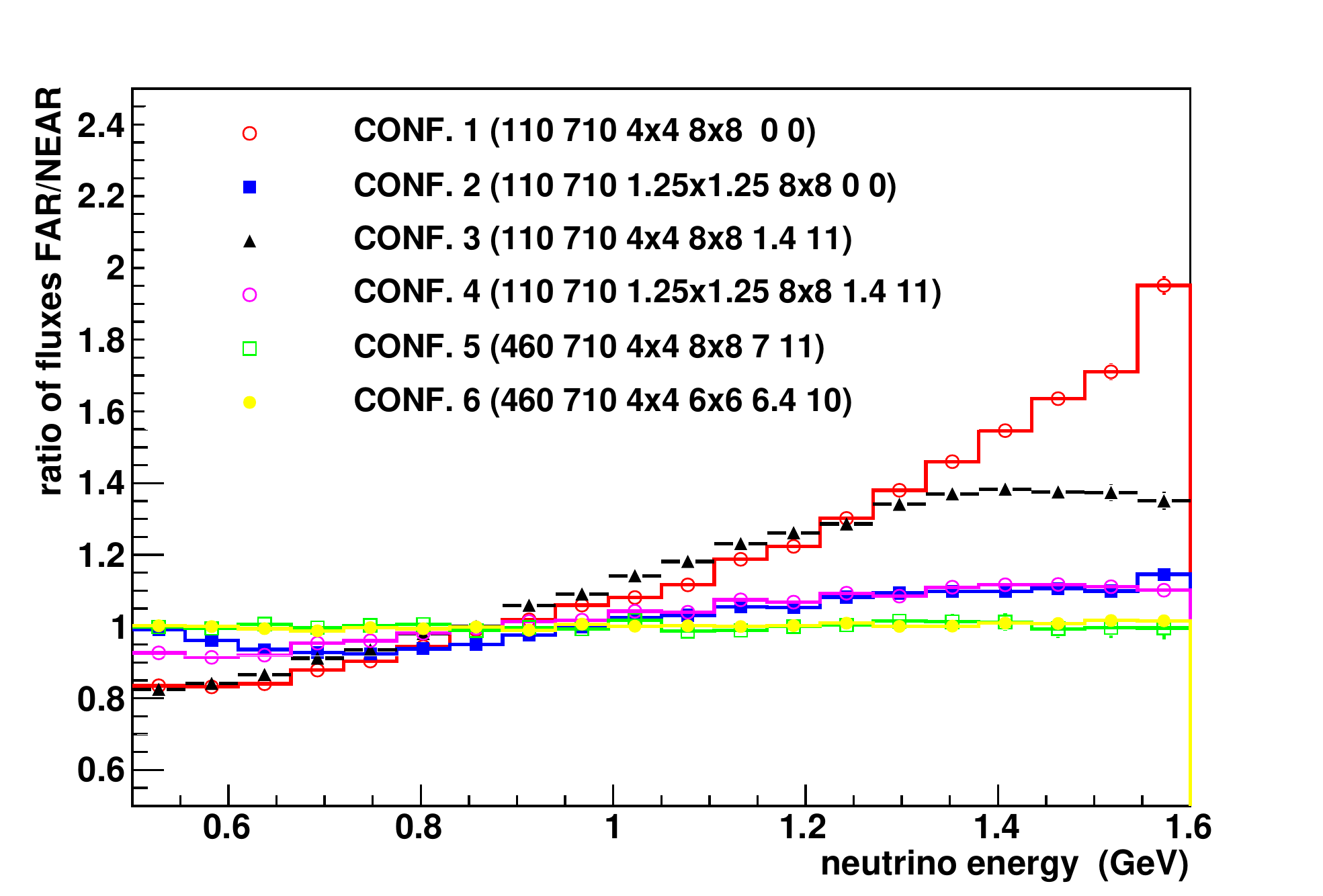}
\caption{Far--to--Near ratios for the six considered configurations using the Sanford--Wang parametrization.}
\label{fig:fig1}
\end{figure}

Configuration 1 (with on--axis detectors and a large Near detector)
produces a FNR increasing with energy as expected from the
considerations presented above, and largely departing from a flat
curve. By restricting the fiducial area in the Near detector
(configuration 2) the FNR flattens out as expected. This behavior is also
confirmed using off--axis detectors (configurations 3 and
4). Configurations with a Near detector at larger baselines (5 and 6)
tend to produce flatter FNRs, as expected.
The different behaviors are more easily visible in Fig.~\ref{fig:fig1} where FNRs, normalized to each other, are compared
 (taking the Sanford--Wang parametrization).
 
The GEANT4 and FLUKA based simulations provide results that are
in agreement at a level between 1\% and 3\% for the configurations
for which only the overlapping regions between the Far and Near detectors are
considered~\cite{nessie-fnal}. A better estimate of the stability of
FNR against hadroproduction uncertaintes has been obtained using
the Sanford-Wang parametrization of pion production data from HARP and
E910 for the target replica. The coefficients $c_i$ in Eq.~\ref{eq:SW} have been
sampled within their uncertainties. The correlations for the uncertainties~\cite{G4BNBflux} have been
properly taken into account using the Cholesky decomposition of the covariance matrix. 
The resulting uncertainties are large (5-7\%)~\cite{nessie-fnal} when taking the
full area of the Near detector at 110 m while they decrease
significantly by restricting to the central region. In particular,
configuration 4, which is realistic from practical considerations, has
an uncertainty ranging from 2\% at low energy decreasing below 1.5-0.5\%
for neutrino energies above 1 GeV. The uncertainty is also quite good
(generally below 0.5\%) for a Near site at 460~m.

In conclusion, using the constraints from HARP/E910 data sets, we
estimate the uncertainties for FNR associated to hadroproduction 
being of the order of 1-2\% for a configuration with the Far detector at
surface and the Near detector at an equivalent off--axis angle and a
fiducial volume tailored to match the acceptance of the Far detector
(Òconfiguration 4Ó). Given also the high available statistics and the
large lever--arm for oscillation studies we consider such a layout with
baselines of 110 m and 710 m as a viable choice.
Of course, ``configuration 4'' is a subset of ``configuration 3'',
which could be that to be used in reality by rearranging the OPERA spectrometers (see next Section). Therefore, given
the possibility for a higher statistics collection and the minor concern about the height of the pit (that has to be anyhow centered at the level
of the beam) the following studies are
based on ``configuration 3''.

\subsection{Spectrometer Design Studies}\label{sec:spect1}

The location of the Near and Far sites corresponds a fundamental issue in the sterile neutrino search. Moreover the two detector systems 
at the two sites have to be as similar as possible. The NESSiE Far spectrometer has to be designed to cope with an aggressive time 
schedule and to largely exploit the acquired experience with the OPERA spectrometers in construction, assembling and 
maintenance~\cite{bopera}. Well known technologies have been considered as well as re--using large parts of existing detectors.
The OPERA spectrometers will begin to be dismantled sometime next year and we foresee to use them all. The relatively low momentum
range of muons from the BNB charged current events suggests to couple together the two OPERA spectrometers, for both
the Far and the Near sites. Their modularity will allow to take 4/7 of the acceptance region (by height) for the Far site and 3/7 for the 
Near one. Each iron slab will be cut at 4/7 in height to reproduce exactly the Far and Near targets.
In this way any inaccuracy either in geometry (the single 5 cm iron slab owns a precision of few mm) or in the material will be 
exactly reproduced in the two detection sites.
The Near NESSiE spectrometer will then be a sacked down clone of the Far one, with identical thickness along the beam but reduced 
transverse size.


The performances of the current geometry with 5 cm thick iron--slabs are evaluated in terms of NC contamination and momentum resolution, and compared to a possible geometry with 2.5 cm slab thickness. 
Using 2.5 cm thick slabs, the fraction of neutrino interactions producing a signal in the RPCs increases for both NC and CC events. 
The CC efficiency and the NC contamination are both larger with respect to the reference 5 cm geometry. 
At the same level of purity the efficiencies in the two geometries are similar. 
No advantage in statistics is obtained requiring the same NC contamination suppression. 

\section{Physics Analysis and Performances}\label{sec:spect2}

We developed sophisticated analyses to determine the sensitivity region that can be explored with an exposure of 
$6.6\times 10^{20}$ p.o.t., corresponding to 3 years of data collection at FNAL--Booster beam.
Our guidelines have been the maximal extension at small values of the mixing angle parameter,
as well as its dependence on systematic effects.

To this aim, three different analyses have been set up, of different complexity:
\begin{itemize}
\item the usual sensitivity plot based on the Feldman\&Cousins technique (see Section V of~\cite{Fel-Cou}), obtained
by adding {\em ad hoc} systematic error evaluations;
\item a full correlation matrix based on the full Monte Carlo simulation including the reconstruction of the simulated data;
\item a new approach based on the profile CLs, similar to that used in the Higgs boson discovery~\cite{atlas-cms}. 
\end{itemize}

Throughout the analyses the detector configuration defined in Table~\ref{tab:NFD} was considered. 

\begin{table}[h]
\centering
\begin{tabular}{|l|c|c|c|c|c|c|}
\hline
 & Fiducial Mass (ton) & Baseline (m)\\
\hline
\hline
Near & 297 & 110\\
\hline
Far & 693 & 710\\
\hline
\end{tabular}
\caption{Fiducial mass and baselines for Near and Far detectors. }
\label{tab:NFD}
\end{table}

Given the relevance of the CCQE component, arising from the convolution of flux and cross--sections,
our analysis makes use of the 
muon momentum as estimator. 
The neutrino energy $E$ is obtained either by the usual formula  in the CCQE approximation
\begin{equation}
E = \frac{E_\mu - m^2_{\mu}/(2M)}{1-(E_\mu -p_\mu\cos\theta)/M},
\end{equation}
\noindent ($M$ being the nucleon mass, and $E_{\mu}$, $p_{\mu}$ the muon energy and momentum, respectively) or via Monte Carlo simulation.

For all analyses the two--flavor neutrino mixing in the approximation of one mass dominance    is considered,
The oscillation probability is  given by the formula:
\begin{equation}\label{eq:2flavour}
P = \sin^{2}(2\theta_{new})\sin^{2}(1.27\ \Delta m^{2}_{new}\ L({\rm km})/E({\rm GeV}))
\end{equation}
where $\Delta m^{2}_{new}$ is the mass splitting between a new heavy--neutrino mass--state and the
heaviest among the three SM neutrinos, and $\theta_{new}$ is the corresponding mixing angle.
As the baseline, $L$, is fixed by the experiment location, the oscillation is naturally driven by the neutrino 
energy, with an {\em amplitude} determined by the mixing parameter.

The disappearance of muon neutrinos due to the presence of an additional sterile state
depends only on terms of the extended PMNS~\cite{pmns} mixing matrix ($U_{\alpha i}$ with $\alpha= e,\mu,\tau$ and $i=1$,\ldots,4)
involving the $\nu_{\mu}$ flavor state and the additional fourth mass eigenstate. In a 3+1 model at Short Baseline (SBL) we have:
\begin{equation}
P(\nu_{\mu}\to\nu_{\mu})_{SBL}^{3+1} = 1 - \left[ 4 \vert U_{\mu 4} \vert^2 (1 - \vert U_{\mu 4} \vert^2)\right] \cdot \sin^2 \frac{\Delta m^2_{41} L}{4E},
\end{equation}
where $4 \vert U_{\mu 4} \vert^2 (1 - \vert U_{\mu 4} \vert^2)$ results as an {\em amplitude}.

In contrast, appearance channels (i.e. $\nu_\mu \to \nu_e$) are driven by
terms that mix up the couplings between the initial and final flavour states and
the sterile state yielding a more complex picture:
\begin{equation}
P(\nu_{\mu}\to\nu_e)_{SBL}^{3+1} = 4 \vert U_{\mu 4}\vert^2 \vert U_{e 4} \vert^2  \sin^2 \frac{\Delta m^2_{41} L}{4E}
\end{equation}
This holds also in extended $3 + n$ models.

It is interesting to notice that the appearance channel is suppressed by two more
powers in $\vert U_{\alpha 4}\vert$. Furthermore, since $\nu_e$ or $\nu_\mu$ appearance
requires $\vert U_{e 4}\vert > 0$ and $\vert U_{\mu 4}\vert > 0$, it should be naturally accompanied by
a corresponding $\nu_e$ and $\nu_\mu$ disappearance. In this sense the disappearance
searches are essential for providing severe constraints  on the models of the theory
(a more extensive discussion on this issue can be found e.g. in Sect. 2 of~\cite{winter}).

It must also be noted that the number of $\nu_e$ neutrinos depends on
the $\nu_e\rightarrow\nu_s$ disappearance and $\nu_\mu\rightarrow\nu_e$ appearance, and, obviously,
from the intrinsic $\nu_e$ contamination in the beam. 
On the other hand, the amount of $\nu_\mu$ neutrinos depends only on the
$\nu_\mu\rightarrow\nu_s$ disappearance and $\nu_e\rightarrow\nu_\mu$ appearance but the latter is
much smaller due to the fact that the $\nu_e$ contamination in $\nu_\mu$ beams is
usually at the percent level. 
Therefore in the $\nu_{\mu}$ disappearance channel the oscillation probabilities in both Near and Far detectors can be measured without any interplay 
of different flavours, i.e. by the same probability amplitude.

\begin{figure}[htbp]
\begin{center}
\includegraphics[scale=0.40]{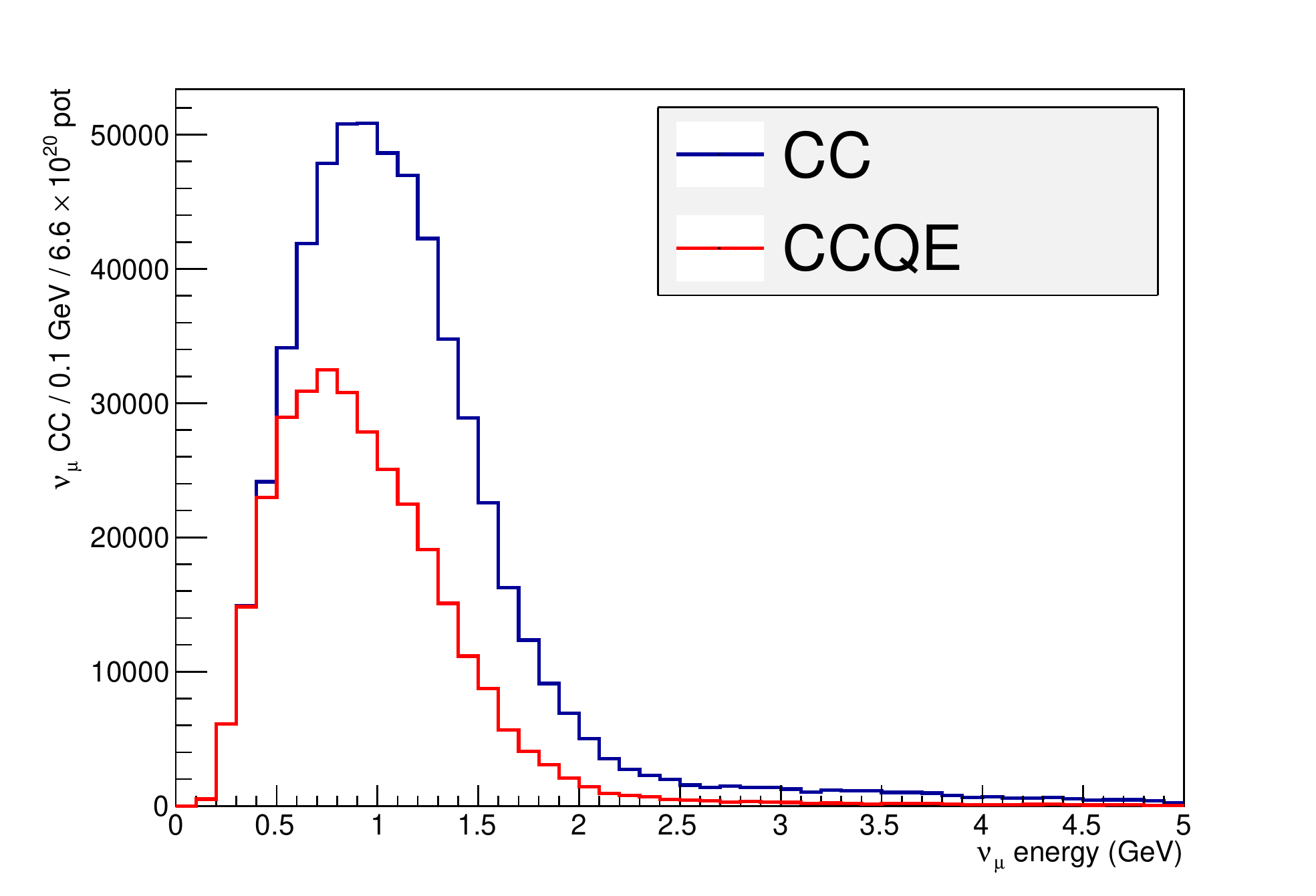}
\includegraphics[scale=0.40]{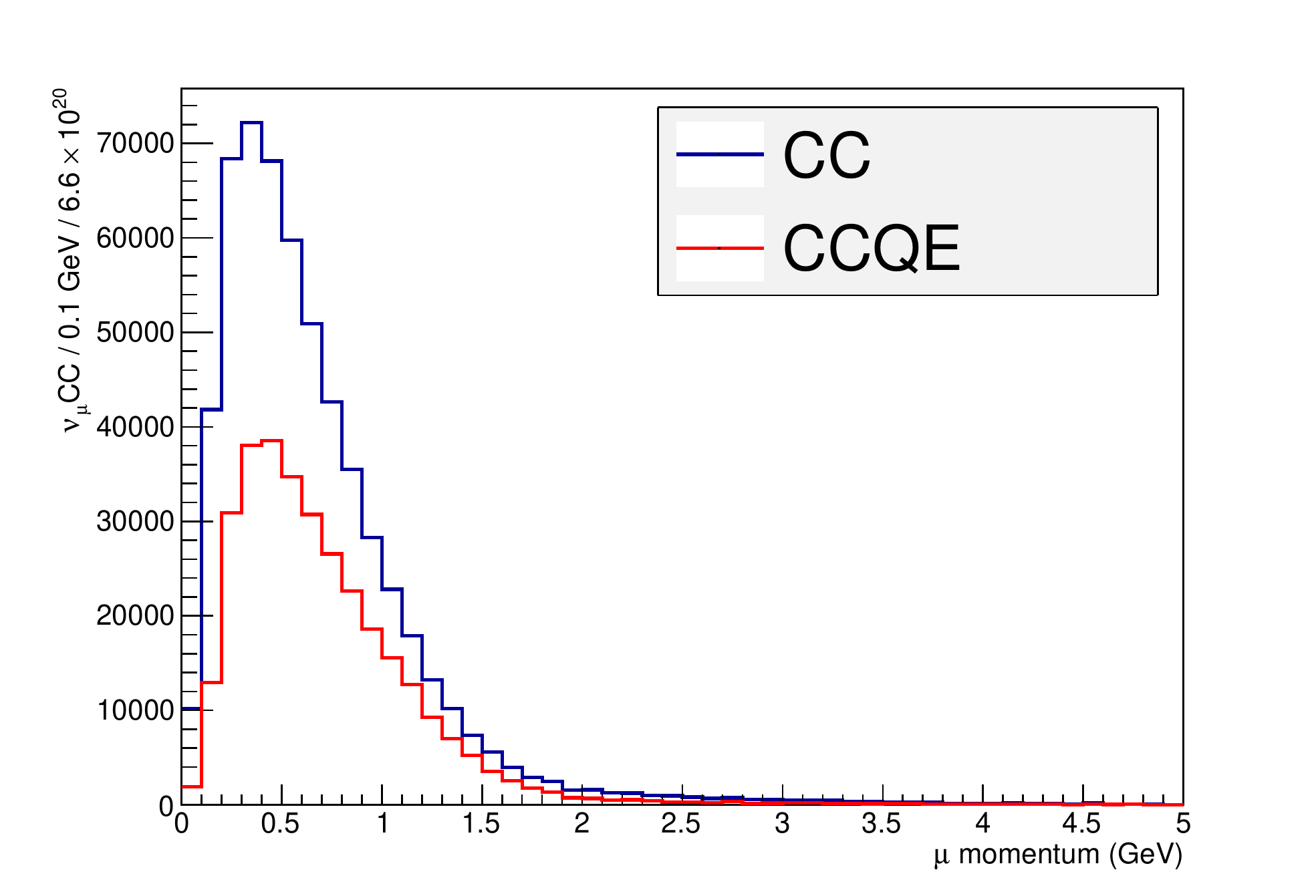}
\caption{The absolute number of \numu CC interactions seen by the Far detector at 710 m, as a function of the
$E_{\nu}$ (left) and the $p_{\mu}$ (right). The sub--sample corresponding to the CCQE component is also shown.}
\label{fig:norma-interac}
\end{center}
\end{figure}

The distributions of events, either in the $E_{\nu}$ or the $p_{\mu}$ variables, normalized to the expected
luminosity in 3 years of data taking at FNAL--Booster, or $6.6\times 10^{20}$ p.o.t., are reported in 
Fig.~\ref{fig:norma-interac}.

\subsection{Sensitivity Analysis}

Three different ways of evaluating the sensitivity region for sterile neutrinos were applied. On top of the standard sensitivity analysis based
on Feldman\&Cousins two more refined analyses have been applied:
\begin{itemize}
\item matrix-correlation,
\item CLs profile likelihood.
\end{itemize}

In the first one we implemented different smearing matrices for two different observables, the muon {\em range} and the {\em number 
of crossed planes}, associated with
the true incoming neutrino energy. These matrices were obtained through a full Monte Carlo simulation.

We studied the sensitivity to the \numu disappearance using two different observables: the range and the number of planes at Near and Far detector, evaluated using GLoBES~\cite{GLoBES}.

The $\nu_{\mu}$ disappearance can be observed either by a deficit of events (normalization) or, alternatively, by a distortion of the observable spectrum (shape) which are affected by systematic uncertainties expressed by the normalization errors matrix and the shape errors matrix,
respectively. The shape errors matrix represents a migration of events across the bins. In this case the uncertainties are associated with changes
not affecting the total number of events and so a depletion of events in some region of the spectrum should be compensated by an enhancement in others. 
Details of the model used for the shape error matrix can be found in~\cite{nessie-fnal}.

By applying the frequentist method the $\chi^{2}$ statistic distribution has been looked at  
in order to calculate the sensitivity to oscillation parameters. The sensitivity computed considering CC and NC events is almost the same as the sensitivity 
obtained with only CC events (see Section 12.2 in~\cite{nessie-fnal}) and therefore NC background events do not affect the result. 

Different cuts on the range and on the number of planes were applied. Sensitivity plots were computed by introducing bin--to--bin correlated systematic uncertainties as expressed in the covariance matrix
by considering either 1\% correlated error in the normalization or 1\% correlated error in the spectrum shape. 

\begin{figure}[htbp]
\centering
{\includegraphics[scale=0.5]{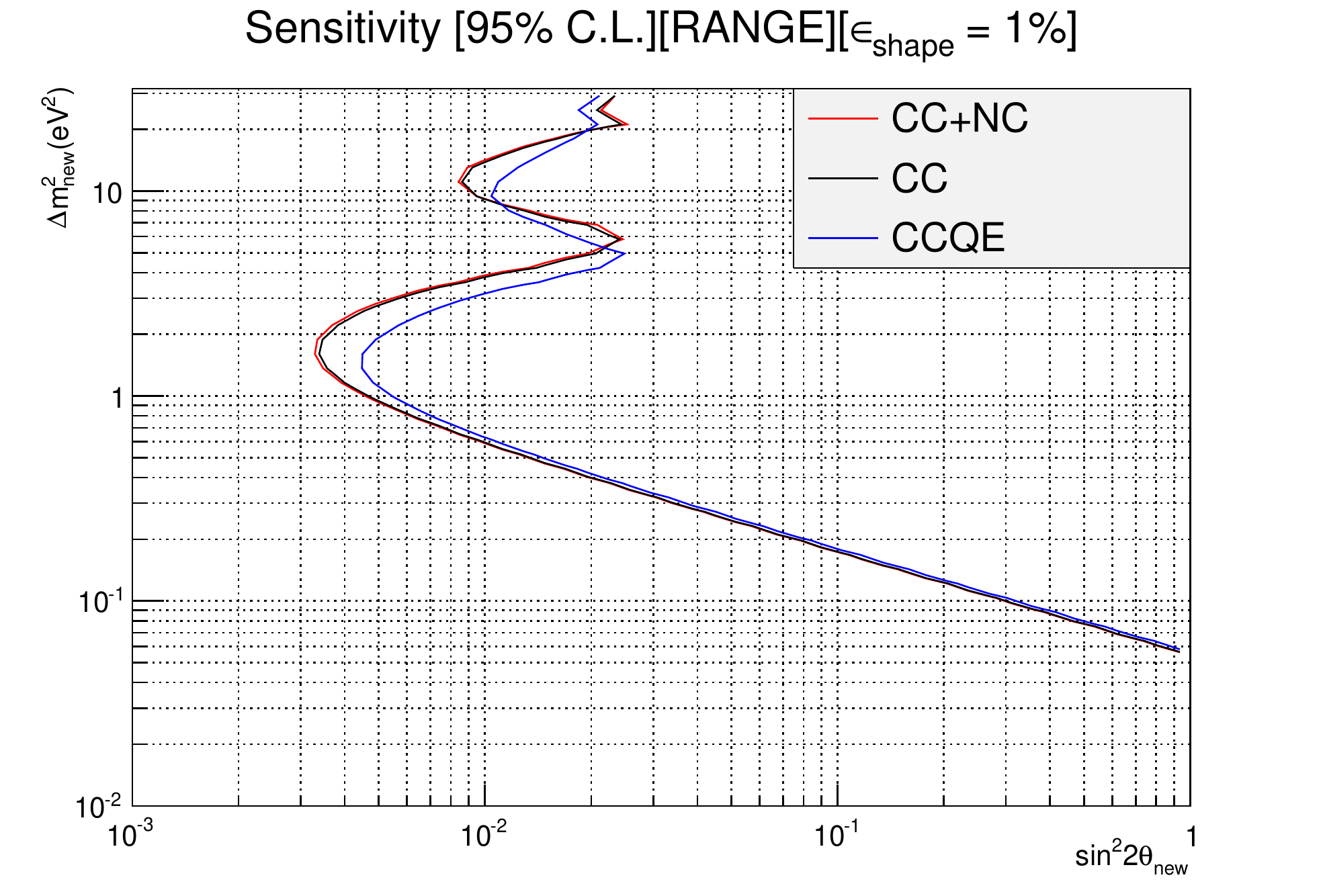}} 
\caption{ 95\% CL sensitivity obtained using range for all  (QE, Res, DIS) CC (black) and CC+NC (red) events and for only CCQE events (blue). In this case we 
considered 1\% bin-to-bin correlated error in the shape.}
\label{fig:sensitivity_plot_all_corrShape}
\end{figure}

As a representative result the sensitivity calculated considering 1\% correlated error in the shape is plotted in 
Fig.~\ref{fig:sensitivity_plot_all_corrShape}. 
It is interesting to outline that the systematic error on normalization affects the sensitivity region only at 
the extreme edges at small values of the mixing angle parameter.

In the profile CLs method we introduce a new test--statistics that depends on a {\em signal--strength} variable. We may observe that,
by looking at Eq.~\ref{eq:2flavour}, for a fixed $\Delta m^{2}_{new}$, the first factor, $\sin^{2}(2\theta_{new})$, acts as an amplification 
quantity of the configuration shape of the estimator being used. Then, we may identify the signal--strength $\mu$ with $\sin^{2}(2\theta_{new})$
and construct the estimator function:
\begin{equation}\label{eq:signal-strength}
f = \frac{1-\mu\cdot\sin^{2}(1.27\ \Delta m^{2}_{new}\ L_{Far}/E)}{1-\mu\cdot\sin^{2}(1.27\ \Delta m^{2}_{new}\ L_{Near}/E)}
\end{equation}

In a simplified way, for each $\Delta m^{2}_{new}$ a sensitivity limit can be obtained from the {\em p--value} of the distribution of the estimator
in Eq.~\ref{eq:signal-strength}, in the assumption of {\em background--only} hypothesis. 

That procedure does not correspond to
computing the exclusion region of a signal, even if it provides confidence for it. The exclusion plot should be obtained by fully
developing the $CL_S$ procedure as outlined above. However, since we are first interested in
exploiting the sensitivity of our experiment to any oscillation pattern not compatible with the standard 3--neutrino scenario,
the above procedure provides insights into such question, and it is fully compatible with the previous two analyses and the usual neutrino analysis
found in literature.

Moreover, following the same attitude, an even more {\em aggressive} procedure can be applied. Since the deconvolution
from $p_{\mu}$ to $E_{\nu}$ introduces a reduction of the information, we investigate whether the more
direct and measurable parameter, $p_{\mu}$, is a valuable one. In such a case Eq.~\ref{eq:signal-strength} becomes:
\begin{equation}\label{eq:signal-strength-mu}
f = \frac{1-\mu\cdot\sin^{2}(1.27\ \Delta m^{2}\ L_{Far}/p_{\mu})}{1-\mu\cdot\sin^{2}(1.27\ \Delta m^{2}\ L_{Near}/p_{\mu})}
\end{equation}

\begin{figure}[h]
\centering
\vskip-1cm
\includegraphics[scale=0.4]{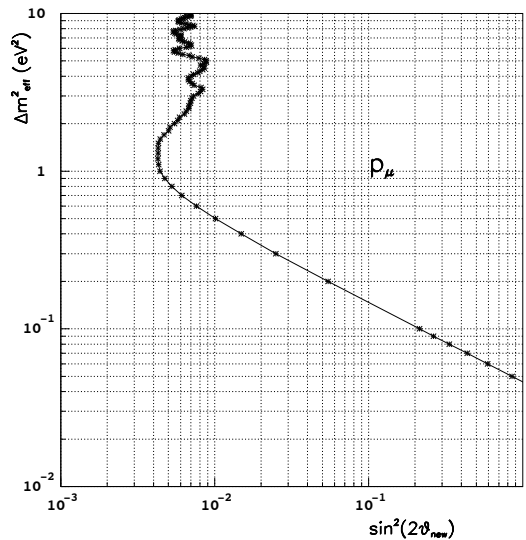}
\vskip-2cm
\caption{The sensitivity plot obtained by computing the modified raster--scan method, in a $CL_S$ framework, and by
using the reconstructed muon momentum as estimator. A conservative cut of $p_{\mu,rec}\ge\ 500$~MeV was applied.}
\label{fig:exclu-pmu}
\end{figure}

The sensitivity plot in Fig.~\ref{fig:exclu-pmu} actually provides an ``effective'' sensitivity limit in the ``effective'' variables
$\Delta m^{2}$  and the reconstructed muon momentum, $p_{\mu,rec}$. We checked that the ``effective'' $\Delta m^{2}$ is simply scaled--off towards higher values, not affecting the mixing angle limit.
This is the best sensitivity result that our experiment can achieve if
the systematics can be limited to 1\% level, as we are confident in. A sensitivity to mixing angles in $\sin^2(2\theta_{new})$ 
below $10^{-2}$ can therefore be obtained in a large region of $\Delta m^{2}$, around the $1\ {\rm eV}^2$ scale.

\section{Conclusions}\label{conclu}
Existing anomalies in the neutrino sector  may hint to the existence of  one or more additional {\em sterile} neutrino families. 
We performed a detailed study of the physics case in order to set a Short--Baseline experiment at the FNAL--Booster neutrino
beam to exploit the measurement of the charged current events. An independent measurement on \numu, complementary to
the already proposed experiments on \nue, is mandatory
to either prove or reject the existence of sterile neutrinos, even in case of null result for \nue.
Moreover, very massive detectors are mandatory to collect a large number of events and therefore improve the disentangling of systematic effects.

The best option in terms of physics reach and funding constraints is provided by two spectrometers based on dipoles iron magnets,
at the Near and Far sites, located at 110 and 710~m from the FNAL--Booster neutrino source, respectively, 
possibly placed behind the proposed LAr detectors. In this way
we will succeed to keep the systematic error at the level of $1\div 2$\% for the measurements of the 
\numu interactions, in particular the measurement of the muon--momentum at the percent level and the identification
of its charge on event--by--event basis, extended to well below 1 GeV.

We plan to perform a full re--use of the OPERA spectrometers that will be dismantled starting at the end of 2015.
Each site at FNAL will host a part of the two coupled OPERA magnets, based on well know technology
allowing to realize "clone" detectors at the Near and Far sites.
The spectrometers will make use of RPC detectors, already 
available, which have demonstrated their robustness and effectiveness.


\section*{Acknowledgements}
\label{sec-ack}

We warmly thank the organizers of ICHEP2014 for the kind invitation to report
about the proposal of the NESSiE Collaboration at the FNAL--Booster neutrino--beam.






\end{document}